\documentclass[preprint,12pt,3p]{elsarticle}

\usepackage{amssymb}
\usepackage{multirow}
\usepackage{graphicx}
\usepackage{subfigure}
\usepackage[ruled,vlined]{algorithm2e}

\journal{}

\begin{document}

\begin{frontmatter}

\title{An Improved Authentication Scheme for BLE Devices with no I/O Capabilities}

\author[label1]{Chandranshu Gupta\corref{cor1}}
\address[label1]{Indian Institute of Technology Jammu, India, 181221}

\cortext[cor1]{I am corresponding author}

\ead{2019pcs0018@iitjammu.ac.in}

\author[label1]{Gaurav Varshney}
\ead{gaurav.varshney@iitjammu.ac.in}


\begin{abstract}
Bluetooth Low Energy (BLE) devices have become very popular because of their Low energy consumption and hence a prolonged battery life. They are being used in smart wearable devices, smart home automation system, beacons and many more areas. BLE uses pairing mechanisms to achieve a level of peer entity authentication as well as encryption. Although, there are a set of pairing mechanisms available but BLE devices having no keyboard or display mechanism (and hence using the Just Works pairing) are still vulnerable. In this paper, we propose and implement, a light-weight digital certificate based authentication mechanism for the BLE devices making use of Just Works model. The proposed model is an add-on to the already existing pairing mechanism and therefore can be easily incorporated in the existing BLE stack. To counter the existing Man-in-The-Middle attack scenario in Just Works pairing (device spoofing), our proposed model allows the client and peripheral to make use of the popular Public Key Infrastructure (PKI) to establish peer entity authentication and a secure cryptographic tunnel for communication. We have also developed a lightweight BLE profiled digital certificate containing the bare minimum fields required for resource constrained devices, which significantly reduces the memory (about 90\% reduction) and energy consumption. We have experimentally evaluated the energy consumption of the device using the proposed pairing mechanism to demonstrate that the model can be easily deployed with less changes to the power requirements of the chips. The model has been formally verified using automatic verification tool for protocol testing.
\end{abstract}


\end{frontmatter}


\section{Introduction}\label{sec1}
\subsection{BLE and Bluetooth}
Bluetooth Low Energy or Bluetooth Smart is a successor of Bluetooth and has captured the market as a preferred medium for communication between low power devices over short distances \cite{Zhang2019}. Bluetooth Low Energy is Bluetooth version 4.0 or above \cite{BLE} and is better than classic Bluetooth technology in terms of latency for a connection, total time to send data, idle time, power and current consumption. This encourage its use in low powered smart devices such as beacons, smart fitness watches, smart home automation systems which need transfer of data over short distances \cite{hussain2017secure}.

BLE is not a direct upgrade to the classic Bluetooth but rather it is a new technology that is focused on Internet of Things (IoT) applications \cite{BLE1,BLE2}. The main difference and similarities between BLE and classic Bluetooth are summarized in the Table 1.

\begin{table}[h]\centering
\footnotesize{
\caption{Difference between BLE and Classic Bluetooth}
{\begin{tabular}{|p{4.0cm}|p{4.7cm}|p{4.2cm}|} \hline
 Specifications & BLE & Bluetooth classic \\ \hline
 RF Frequency band & 2.4 GHz & 2.4 GHz  \\
 Current consumption & $<$15 mA & $<$30 mA \\
 Payload & 251 B & 1021 B \\
 Slaves & unlimited & 7 \\
 Usage & Short burst data transmission & Continuous data streaming \\
 Security & 128-bit & 56/128-bit \\
 Power consumption & 0.01 to 0.50 W & 1 W \\ 
 Data rate & 125 Kb/s to 2 Mb/s & 1 Mb/s to 3 Mb/s \\ \hline
\end{tabular}}}
\label{IoT}
\end{table}

\subsection{BLE stack and pairing mechanism}
\begin{figure}[ht]
    \centering
    \includegraphics[keepaspectratio,height=9cm,width=9cm]{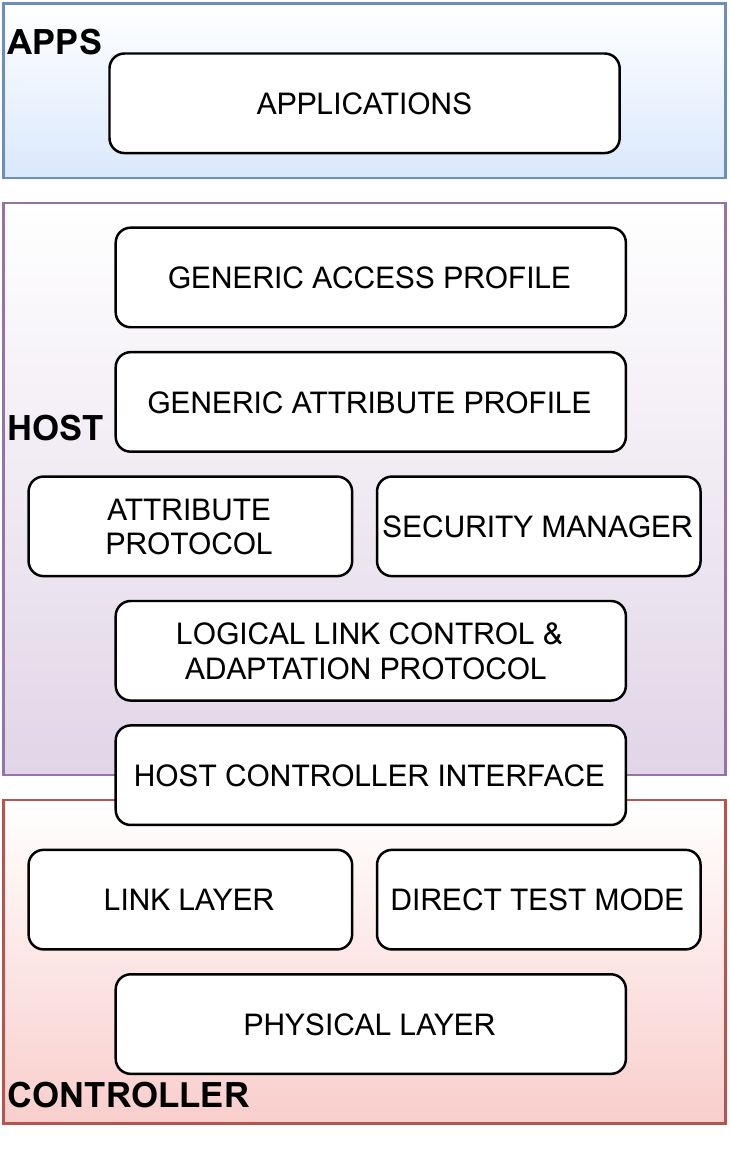}
    \caption{BLE stack}
    \label{BLEstack}
\end{figure}

Figure \ref{BLEstack} depicts the BLE stack, It has three components: \texttt{Controller}, \texttt{Host} and the \texttt{Application}. Controller and Host are interfaced using \texttt{Host controller Interface (HCI)}. The Controller consists of \texttt{Physical layer, Direct Test Mode and Link Layer} \cite{BLE3}. The Host includes \texttt{Logical Link Control and Adaptation Protocol (L2CAP), the Attribute Protocol (ATT), the Generic Attribute Profile (GATT), the Security Manager (SM) and the Generic Access Profile (GAP)}. 

The physical layer is responsible for transmitting and receiving data over Radio waves. The LE radio operates in 2.4 GHz ISM band \cite{BLE5}.The Link layer interfaces directly with the physical layer \cite{BLE6}. It defines certain roles:
\begin{itemize}
\small
    \item \texttt{Advertiser/Scanner} (Initiator)
    \item \texttt{Slave/Master or Peripheral/Client}
    \item \texttt{Broadcaster/Observer}
\end{itemize}
All these roles are defined in five states: \texttt{Scanning, Advertising, Standby, Initiating and Connected}.

L2CAP layer performs fragmentation and de-fragmentation of data and multiplexing and de-multiplexing of channels. It interfaces with upper layer protocols and provides error checking, window based flow control scheme and re-transmission \cite{BLE,BLE6}. Above the L2CAP layer is Attribute Protocol (ATT) which defines how the server advertises its data to the client device and also defines how the data is structured within the server. The data is structured in the form of \texttt{Attributes}. ATT defines two kinds of roles which includes a \texttt{BLE Server and a BLE Client}. Devices in BLE Server mode (such as smart fitness trackers) generally exposes the attributes which can be discovered, read and written on by a BLE client device (such as Smart phones). The attribute structure consists of: \texttt{Attribute Type (defined by UUID), Attribute Handle (16-Bit Unique Identifier), Attribute Value and Permissions} \cite{BLE6}.

Generic Attribute Profile (GATT) built over ATT, specifies the way the data should be sent or received between the devices. GATT takes the same roles as ATT but these roles are not set per device but are determined by the transaction between the devices. The operating system interacting with BLE devices use GATT concepts which define various profiles which segregate a specific set of services. For example there can be a set of health care profiles such as \texttt{Health Thermometer Profile} which might be of interest to a BLE based Thermometer device while there can be a set of sport profiles such as \texttt{Heart Rate Profile} which may interest a device manufacturer dealing with BLE based smart fitness devices. Each BLE profile define a set of \texttt{Services} which are composed of \texttt{Characteristics}. Each characteristic stores a \texttt{Value and Properties} which relates to its operation as well as some \texttt{Descriptors} which stores meta data information about that characteristics. These services and characteristics store and define the way data will be accessed by another BLE device \cite{willingham2018testing}. 

Generic Access Profile (GAP) is present at the top of the Host \cite{wu2020blesa}. It defines two main aspects: \texttt{Device discovery} (through scanning and advertising) and \texttt{Establishing and Managing Secure Connections} between the devices. GAP offers four different roles which describes the type of device:
\begin{itemize}
\small
    \item \texttt{Broadcaster} (only sends the advertisements)
    \item \texttt{Observer} (only receives/reads the advertisements)
    \item \texttt{Peripheral} (advertises and allows connection establishment)
    \item \texttt{Central} (looks for advertisements and initiates the connection)
\end{itemize}

\begin{figure*}[htbp]
    \subfigure[LE legacy pairing]
    {\includegraphics[width = 3.5 in, height=5 in]{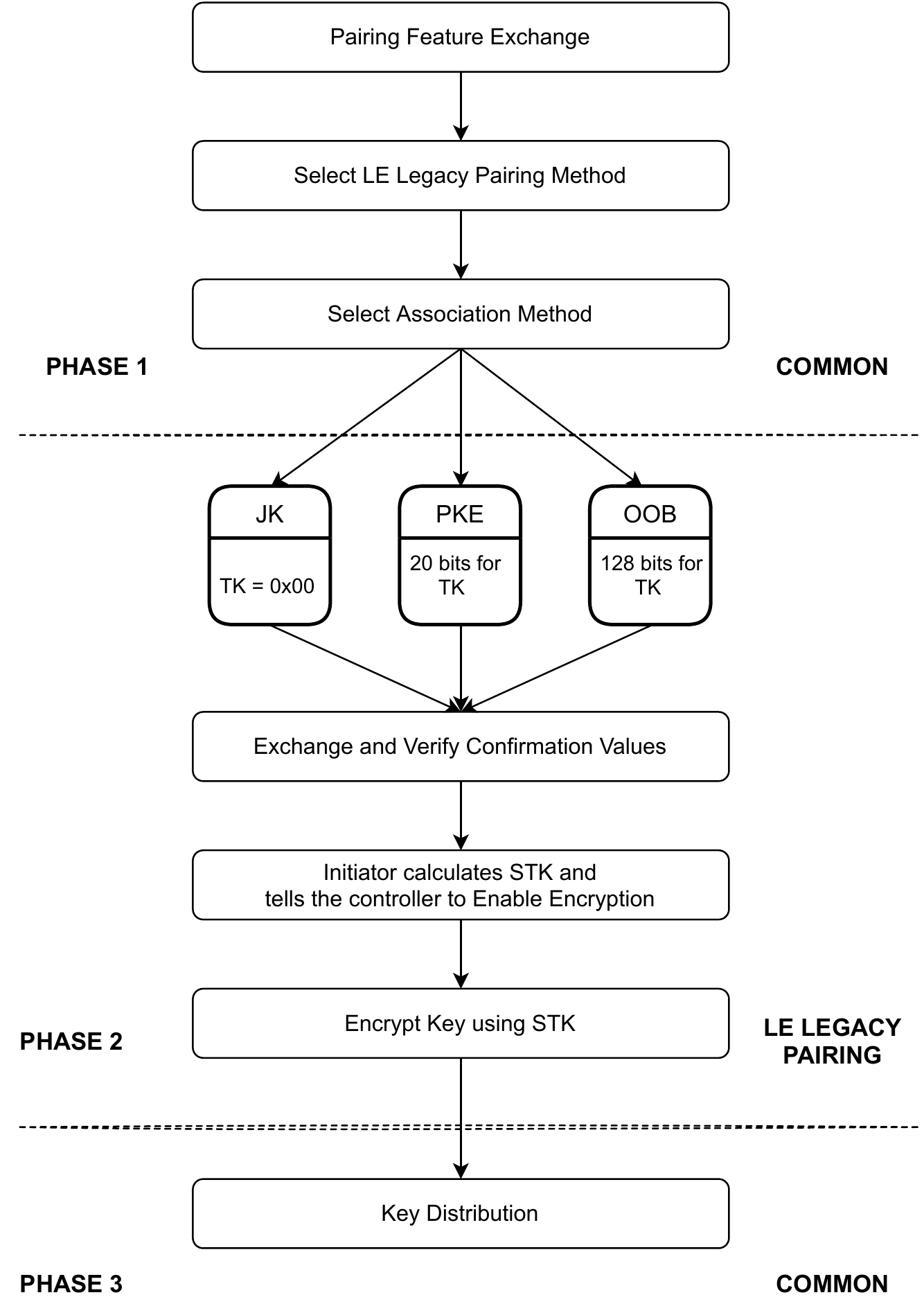}}
    \subfigure[LE Secure connections]
    {\includegraphics[width = 3.5 in, height=5 in]{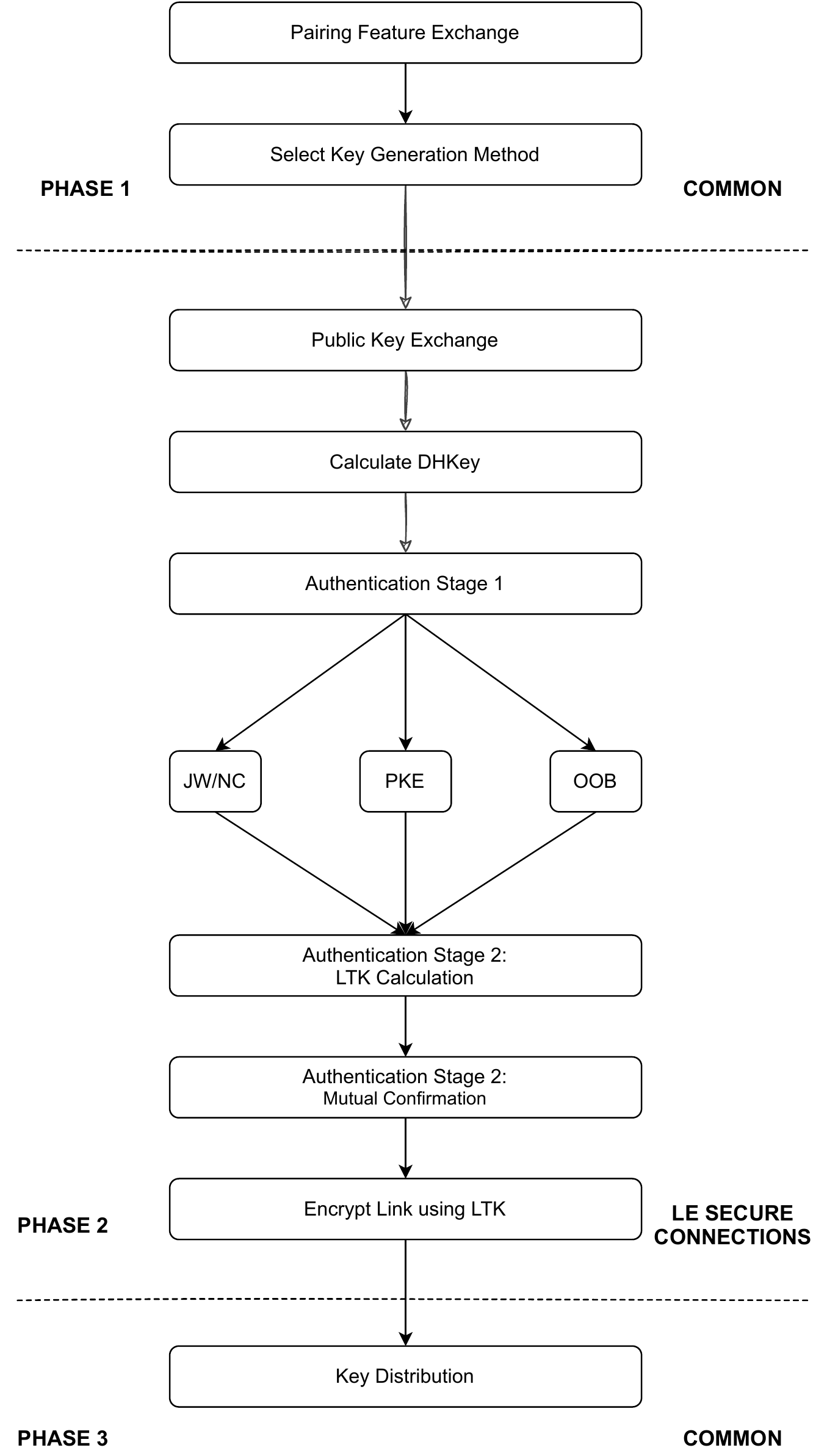}}
    \caption{LE legacy pairing and LE secure connections phases}
    \label{phases}
\end{figure*}

Security Manager (SM) present above the L2CAP protocol, defines the protocol to perform and manage all the security related tasks, that includes: \texttt{Pairing, Bonding, Authentication and Encryption} \cite{BLE7}. \cite{Zhang2019} BLE pairing differs from the classic Bluetooth pairing. BLE pairing comprises of three phases that includes : \texttt{Pairing Feature Exchange}, \texttt{Short Term Key (STK) generation} using Temporary key (TK) for LE legacy pairing or \texttt{Long Term key (LTK) generation} for LE secure Connections and \texttt{Transport specific key distribution}. During the pairing feature exchange phase the devices share their Input/Output (I/O) capabilities, authentication requirements through \texttt{pairing request} and \texttt{pairing response packets} 7 bytes each. The pairing request and response message contains one byte each for \texttt{code description, IO capabilities, out-of-band data flag, specification of authentication requirements, maximum key size, initiator key distribution and responder key distribution specifications}. Figure \ref{phases}. shows the LE legacy pairing and LE secure connections phases \cite{BLE3}.

\begin{figure*}[htbp]
    \centering
    \includegraphics[width=0.75\textwidth]{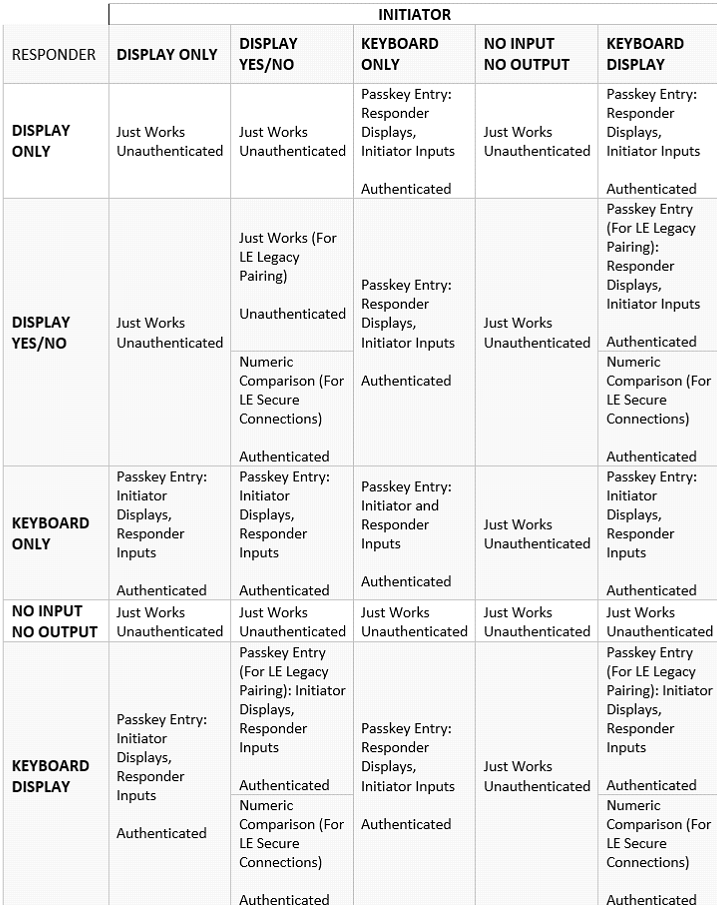}
    \caption{Mapping of I/O capabilities to key generation method used}
    \label{IOcapabilities}
\end{figure*}

Both the devices select a key generation method once the pairing request and pairing response messages are shared with the information required for establishing a connection. The key generation method depends on the IO capabilities and whether LE secure connections or LE legacy pairing is dependent on the Authentication requirements flag \cite{BLE4}. The set of key generation methods differs for LE legacy pairing and for LE secure connections (for \texttt{BLE 4.2} devices). The key generation methods used in LE legacy pairing (for \texttt{BLE 4.0} and \texttt{BLE 4.1} connections) includes: \texttt{Just Works, Passkey and Out-OF-Band (OOB)}. For LE secure connections \texttt{Numeric Comparison} method is also available. Figure \ref{IOcapabilities} shows the Mapping of I/O capabilities to key generation method used.


\begin{enumerate}
\small{
    \item Just Works: In this association model when LE legacy pairing is chosen, STK is generated using TK as one of its inputs \cite{BLE3}. The TK used in Just Works model consists of \texttt{all zeroes} and hence has 0 bits of entropy. Therefore this method doesn't provide MITM protection. In case of LE secure connections both the devices initially sends their \texttt{public keys} to each other and then computes a shared secret using \texttt{Elliptic Curve Diffie Hellman (ECDH)} algorithm, then a confirm value is generated and sent to the other device along with a Nonce. The receiver then calculates the confirm value using the public keys and nonce. The value is then compared with the received confirm value. Even this solution doesn't provide MITM protection as there is no human intervention.
    \item Passkey: In this association model, a \texttt{6 digit passkey} is generated on one device and entered on the other device. The passkey is never transferred over the air and hence provides MITM protection. The problem here is that this method makes use of a small TK, which only provides 20-bits of entropy \cite{BLE8,BLE9}.
    \item OOB: Here, \texttt{128-bit value} is passed out of band between the two devices. It makes use of technology other than BLE (\texttt{such as NFC}) for providing authentication \cite{BLE3}.
    \item Numeric comparison: This is only available in LE secure connection and follows the same procedure as Just works but adds one more step at the end \cite{BLE4}. Both the devices calculates a \texttt{6 digit number} using the public keys and nonce and displays it on the screen. The user verifies whether the number displayed on both devices are same or not and hence authentication is done \cite{BLE9, BLE4} }
\end{enumerate}

\subsection{Research Problem and Motivation}
Although, BLE specification does provide four associative models for performing pairing mechanism for data encryption/authentication and only Numeric comparison method is considered to be secure against MITM attack. However, most BLE devices are resource constrained and therefore cannot make use of Numeric comparison method. Hence, these devices must make use of other three models, which are deemed insecure \cite{ryan2013bluetooth,qu2016assessing}. The devices that do not possess both a keyboard and a display are forced to make use of Just Works method, which is most vulnerable and attack prone. In reality, pairing is considered optional when connecting two BLE devices. Most of the devices turn this feature off to gain battery life but in-turn has to sacrifice security. Research by \cite{zhang2017security} proved that most of BLE wearable devices are vulnerable. An attacker can sniff the unencrypted data and can collect and control sensitive information easily. On an unencrypted an unauthenticated channel device spoofing can happen and BLE smart devices in the proximity can be controlled by a hacker. 

This is a major reason why Bluetooth Low Energy devices cannot be relied for communication of sensitive information in a security critical environment. Though there are other major problems including device tracking, passive eaves dropping but MITM is a major one to be addressed to ensure peer entity authentication and confidentiality. Even if someone use BLE Address as a way to authenticate devices that are using Just Works pairing (devices with no display and keyboard) an attacker in proximity can spoof the BLE hardware address on another BLE chip to evade such authentication. All the key generation schemes which get utilized during phase 2 (other than Numeric comparison) are not a secure choice. The situation gets more troublesome when Just Works method is to be used because:

\begin{itemize}
\small
    \item The secure pairing methods such as Passkey and Numeric Comparison needs either a keyboard and/or a display on the BLE device. Devices such as \texttt{beacons, smart bulbs} and other home automation devices, which do not have a display, and/or a keyboard and hence cannot utilize the security provided by Passkey or Numeric Comparison method.
    \item Device spoofing can result in MITM attacks where a malicious device can sit in between the BLE Central and BLE Peripheral and can do the passive eavesdropping \cite{ryan2013bluetooth} or can deceive other devices and communicate as a trusted device on the network. In such an attack a malicious device can impersonate as a legitimate device to both the devices in communication and can intercept all data being sent between the two devices and inject or modify data as and when needed.
\end{itemize}

Therefore, there is a need to develop a secure authentication scheme for BLE communication, especially when Just Works is the only possible model that can be used because of no I/O capabilities of the BLE devices involved in the communication. Hence, we have proposed a digital certificate based authentication scheme to solve the issue of MITM attack when Just Works method is to be used.

\subsection{Major Contributions}

To solve the problem of MITM through device impersonation (device spoofing) we have proposed the use of light weight PKI based authentication between BLE central and BLE peripheral devices that makes use of Just Works model. The proposed solution is a supplement to an already existing pairing mechanism. As the PKI based communication is already tested to be secure over the web, we believe leveraging its benefits in the BLE communication will help achieve the needed level of security which BLE community may be looking for \cite{forsby2017lightweight}. We have even proposed a lightweight \texttt{BLE profiled Certificate} which only makes use of the required certificate fields as compared to a \texttt{X.509 Certificate} and hence reducing the memory usage. This certificate has been modelled by keeping \texttt{DTLS IoT Profile Certificate} in mind \cite{rfc7925}.


\section{Related Work}
The most challenging aspect of security space is the key exchange mechanism. The key exchange mechanism used by LE legacy pairing mechanism is proven insecure against passive eavesdropper and MITM attack \cite{ryan2013bluetooth}. Rosa\cite{rosa2013bypassing} proved that in LE legacy pairing the commitment value is flawed and hence the attacker can connect to the peripheral device using passkey method without even knowing the six digit key. Zegeye \cite{zegeye2015exploiting} in their research exposed the BLE TK used in calculating STK with the help of brute force attack. William and Melamed \cite{oliff2017evaluating,melamed2018active} presented both software based and hardware based attacks between a BLE device and its smartphone application to exploit the vulnerabilities of the spoofing and MITM attack. Research done by Haattaja and Toivanen \cite{haataja2010two} on \texttt{Bluetooth Secure Simple Pairing (SSP)} security proved that MITM attack is possible on SSP Just works pairing mechanism. This attack is therefore also applicable to LE secure connection Just Works as well. Also their study showed that the attacker can sniff and tamper the packets sent during pairing feature exchange and can force the user to perform insecure pairing mechanism. This means an attacker can force the user to choose insecure Just works method instead of a secure Numeric comparison method.

To provide solution to above problems, Researchers have come up with few solutions with different approaches. Qiaoyang Zhang \cite{zhang2019developing} provided a new security framework where a client and server can establish secure keys using a QR code and use them to secure the BLE channel for message transmission. Gong \cite{gong2018securing} proposed a new encryption algorithm for IoT devices that help in improving the transmission success rate. Yue Zhang et al. \cite{zhang2020breaking} built a prototype for the SCO mode so that the SCO initiation can be properly handled by initiator. Unlike these works, the solution provided by in this paper is aimed at a lightweight authentication mechanism for the BLE devices making use of Just Works association model. Since, none of the existing proposals solve the authentication problem in Just Works model, we have proposed a mechanism that makes use of digital certificates (PKI) for establishing secure session keys between both the client and server for establishing a encrypted communication tunnel. Moreover, the framework proposed by us does not assume any change in BLE architecture and is a supplement to already existing pairing mechanism.

\section{An Improved Authentication Scheme for BLE Devices with no I/O Capabilities}
We propose a digital signature based secure authentication scheme which can be utilized during BLE pairing and authentication for preventing device spoofing and MITM attacks. The realization of the proposed scheme requires understanding of three important phases of its use.
\begin{enumerate}
\small
    \item \texttt{BLE Chip Certificate Generation Phase}
    \item \texttt{BLE Pairing and Authentication Phase}
    \item \texttt{BLE Certificate Revocations and Updates}
\end{enumerate}
\subsection{BLE Chip Certificate Generation Phase}
To realize our approach in practice we need the help of BLE chip manufacturers. The \texttt{BLE Chip Manufactures (BLECM)} must follow a standard and predefined algorithm (ex: \texttt{NIST Curve P256 ECC} algorithm) to generate private and public keys to be used by an asymmetric encryption algorithm for digital signature activities once the BLE chip is shipped. In our case, we use ECC algorithm-generated key pair. 
\begin{enumerate}
    \item The BLE device generates private and public key pair based on the predefined ECC parameters once the device is booted for the first time. It then sends its public key to the \texttt{Original Equipment Manufacturer (OEM)}/BLECM. Also the private key of the device is securely stored such that it can't be read by any third party.
    \item BLECM then digitally sign the information of the Chip using its private key and sends the public key, details of the BLE chip such as hardware address/identifier and the details of the BLE chip manufacturer to \texttt{BLE Certification Authority (BLECA)}. The BLECA after verifying the digitally signed certificate issuing request from a registered BLECM issues a public key certificate for the manufactured chip to the BLECM. Certificate's Subject can be identified through BLE hardware address/identifier or a unique chip manufacturer code for the device.
    \item The certificate then gets stored in the BLECM database or inside the BLE device itself. BLECM maintains a secure and reliable service to store and access certificates belonging to the BLE chips manufactured by it. In our case we have considered BLECA as a BLECM for an easy understanding.
\end{enumerate}

\begin{figure}[h]
    \centering
    \includegraphics[keepaspectratio,width=0.7\textwidth]{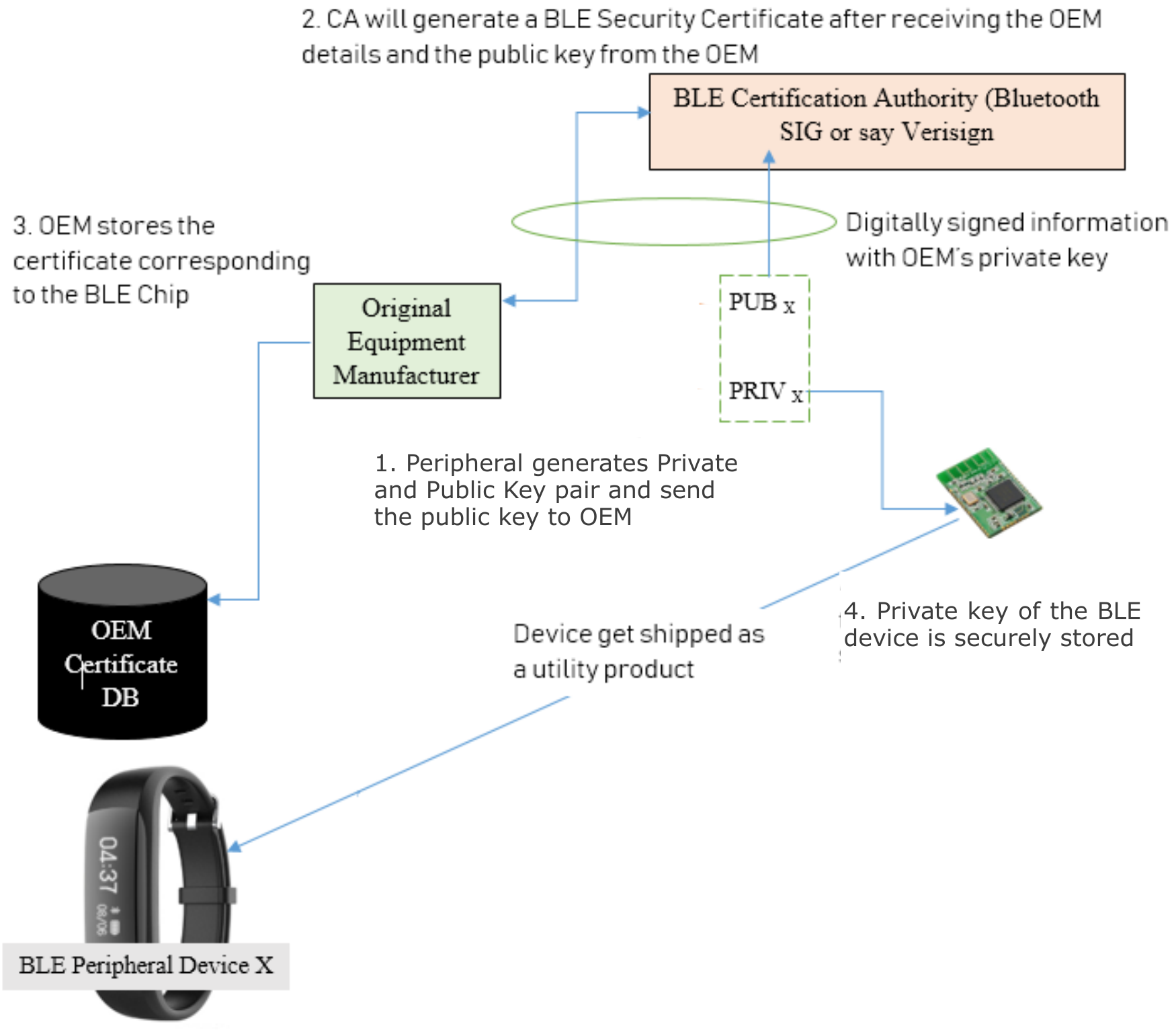}
    \caption{BLE chip specific certificate generation phase}
    \label{fig1}
\end{figure}   

Once above process completes the BLE chip gets shipped to other companies who eventually manufacture the final product to the customer. The product can be a BLE enabled beacon, a BLE enabled smart fitness wrist watch or a BLE enabled smart door lock etc.

\subsection{BLE Pairing/Authentication Phase}

\begin{figure}[htbp]
    \centering
    \includegraphics[keepaspectratio,width=0.9\textwidth]{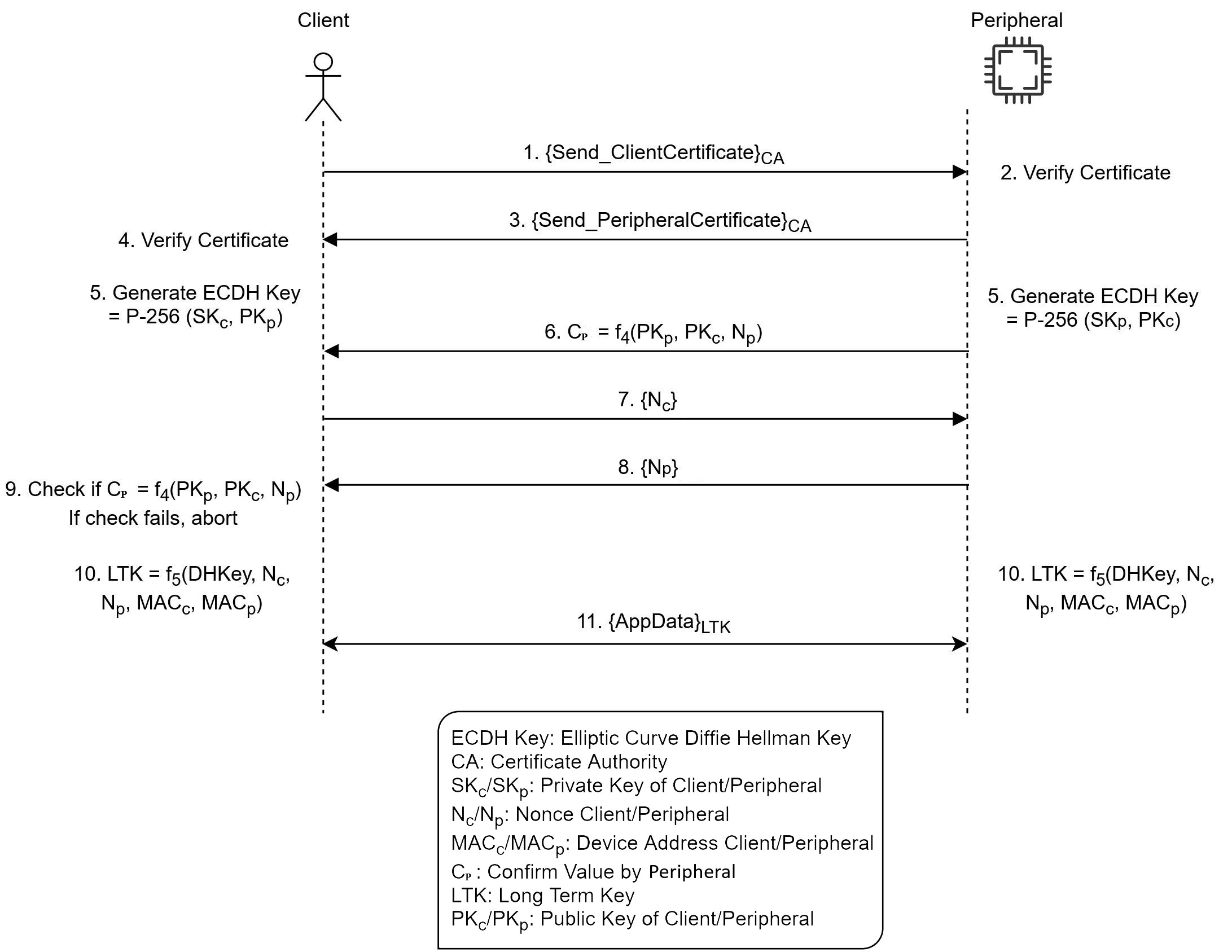}
    \caption{BLE authentication and key generation phase}
    \label{fig5}
\end{figure}

In BLE authentication phase, the pairing feature exchange happens the same way as in the existing BLE pairing. The BLE central and peripheral exchange the 7 bytes data but include a predefined bit as `1' in the `3' bit reserved field of the \texttt{Authentication Requirement Octet}. When one of the predefined reserved bit is set to `1', the devices need to perform the proposed authentication mechanism. The details of the steps performed are explained below:


\begin{enumerate}
\small
    \item As shown in Figure \ref{fig5} the client/central device will initiate the request for the certificate and it will send its own public key or may send its certificate containing the public key to the peripheral, to which the peripheral device replies with its certificate. Since both the devices trust the \texttt{Root Certificate Authority} or BLECA and both of them have public keys of the BLECA, they can use this to verify the public key certificates and hence can trust the public keys. After this both the devices establish a shared secret (ECDH Key) using ECDH key exchange mechanism over NIST P-256 curve and this marks the end of public key exchange.
    \item After this, both the devices selects a pseudo-random \texttt{128-bit nonce} which will devise the LTK. Following this the peripheral device computes the \texttt{confirm value} using the two public keys and its own nonce value. This confirm value is computed as a \texttt{one-way function f4}, which is generated using \texttt{AES-CMAC} with the nonce generated by the peripheral used as the key.
    \item  The client/central then sends its own nonce to the peripheral. In response the peripheral shares its nonce value to the client. The client device then checks whether the confirm value is valid or not since it has received everything to re-calculate the confirm value. If the confirm value is not same, then device can't proceed further and the connection will be terminated.
    \item Finally both the devices computes the LTK, using the shared secret (ECDH Key), two random nonce values generated earlier and the device hardware addresses of both the client and peripheral. Here LTK is generated using the \texttt{function f5}, which is a key generation function and it makes use of AES-CMAC algorithm with ECDH Key as the key. This LTK is then used for encrypting the link.
\end{enumerate}

During the authentication and key exchange process, confirm values are exchanged. These confirm values are computed using the confirm value generation function f4. This confirm value generation function makes use of the MAC function AES-CMAC\textsubscript{x} , with a 128-bit Key X. 

The inputs to function f4 are: \texttt{PK\textsubscript{c} is 256 bits, PK\textsubscript{p} is 256 bits, N\textsubscript{p} is 128 bits}
\begin{equation}
    \texttt{C\textsubscript{p} = f4(PK\textsubscript{c}, PK\textsubscript{p}, N\textsubscript{p})
= AES-CMAC\textsubscript{N\textsubscript{c}} (PK\textsubscript{c},PK\textsubscript{p},N\textsubscript{p})}
\end{equation}

The function f5 is used to generate the LTK. The definition of this key generation function makes use of the MAC function AES-CMAC\textsubscript{T} with a 128-bit key T. The inputs to function f5 are: \texttt{ECDHkey is 256 bits, N\textsubscript{c} is 128 bits, N\textsubscript{p} is 128 bits, MAC\textsubscript{c} is 56 bits, MAC\textsubscript{p} is 56 bits}

The key (T) is computed as follows:
\begin{equation}
 \texttt {T = AES-CMAC\textsubscript{SALT}(ECDHkey)}   
\end{equation}

SALT is the 128-bit value: \texttt {0x6C88 8391 AAF5 A538 6037 0BDB 5A60 83BE}
\begin{equation}
    \texttt {LTK = f5(N\textsubscript{c}, N\textsubscript{p}, MAC\textsubscript{c}, MAC\textsubscript{p})
= AES-CMAC\textsubscript{T}(N\textsubscript{c}, N\textsubscript{p}, MAC\textsubscript{c}, MAC\textsubscript{p}) }
\end{equation}

\begin{algorithm}[H]
\SetAlgoLined
Output: {LTK} \\
Variables:
 G (Base point or Generator of the subgroup), d\textsubscript{c} (Client's Private Key), d\textsubscript{p} (Peripheral's Private Key), Client\_Certificate (Contains Client's Public Key `H\textsubscript{c}=d\textsubscript{c}G'), Peripheral\_Certificate (Contains Peripheral's Public Key `H\textsubscript{p}=d\textsubscript{p}G'), ECDHKey (Shared Secret between Client and Peripheral), MAC\textsubscript{c} (Device Public address of Client), MAC\textsubscript{p} (Device public address of Peripheral), N\textsubscript{c} (Client's 128 bit Nonce), N\textsubscript{p} (Peripheral's 128 bit Nonce), C\textsubscript{p} (Confirm value by Peripheral), LTK (Long Term Key)\\
 \eIf {verify(Client\_Certificate \&\& Peripheral\_Certificate) == Success}{
  Generate ECDHKey = d\textsubscript{c}H\textsubscript{p}= d\textsubscript{p}H\textsubscript{c}\;
  (N\textsubscript{c},N\textsubscript{p}) = Generate\_PRNG\;
  C\textsubscript{p} = AES-CMAC\textsubscript{N\textsubscript{p}}(H\textsubscript{c}, H\textsubscript{p})\;
  C\textsubscript{p} $->$ Client\;
  N\textsubscript{c} $->$ Peripheral\;
  N\textsubscript{p} $->$ Client\;
  Client Calculate C\textsubscript{p}\;
  \eIf{(Calculated C\textsubscript{p} == Received C\textsubscript{p})}{
   Generate LTK = AES-CMAC\textsubscript{ECDHKey}(N\textsubscript{c}, N\textsubscript{p}, MAC\textsubscript{c}, MAC\textsubscript{p})\;
   }{
   Abort: C\textsubscript{p} didn't match \;
  }
  }{Invalid Certificate, abort\;}
 \caption{BLE Authentication Sequence}
\end{algorithm}


\subsection{BLE Certificate Revocations and Updates}  Whenever the Private Key of the device is compromised and the BLE certificates need to be revoked the private keys in the devices need to be updated to publish the new BLE certificates. To do this the BLE company who manufactured the device can push this into firmware updates that they do periodically. In case of update to a peripheral device the firmware update App (firmware updates through company App installed on a BLE central device such as phone) can communicate with the BLE device and set a specific (reserved) bit in the pairing feature exchange process. Once the connection establishes it look for the first message encrypted with the Private Key of the device (hard coded in the device and non readable or writable by third parties) having new BLE certificate key established generated by the BLE device manufacture for that device and the current time stamp. Once the time stamp is in bounds the peripheral device updates the storage of digital certificate's private key with the new key. If the received information from the BLE central App does not decrypts well or the time stamp is old, the peripheral gets disconnected by itself. The update is a tedious process and in this scenario the process of generation and distribution of the private keys is in the hands of the BLE device manufacturer. This is an initial proposal but we are currently studying more secure ways to take care of certificate revocations and it is a matter which needs more work in the future. 

\subsection{Light weight certificate for BLE devices}
We propose a lightweight certificate for BLE devices. In our proposal, we have kept in mind the guidelines mentioned in \texttt{DTLS} profile for \texttt{IoT} standard.
\begin{itemize}
\small
    \item \texttt{Version} : It is a 1 Byte field and whenever there is a version update, it can be incremented. We have kept it as 1.
    \item \texttt{Serial number} : It is a 6 Byte unique field containing the MAC address of the BLE device and therefore will help to identify the device. It will also be used to identify the issuer of the certificate the device manufacturer. During the secure Authentication the static MAC of the device will be used in cases where devices are switched to work on Random MAC. 
    \item \texttt{Subject Public Key Info} : It is a 32 Byte field that contains the Public Key. Here we restrict the algorithm to use 256 bits ECC keys from the curve prime256v1.
    \item \texttt{Signature} : It is a 64 byte field. This field is used to verify the certificate. In our case since we are using ECDSA (RFC5480) with SHA 256.
    
\end{itemize}
Here we have kept the Digital Signature Algorithm and Public Key algorithm as fixed (after experimenting that they are most energy efficient) and hence not allocated a field in the Certificate. In future the same can be added by the vendors. Table \ref{table2} below compares the size of different fields in a X.509 certificate and BLE profiled certificate.
\begin{table}[h]\centering
\caption{Comparison of the Size of X.509 v3 vs proposed BLE Certificate}
{\begin{tabular}{|p{5.7cm}|p{2.4cm}|p{2.4cm}|} \hline
 & \multicolumn{2}{c|}{Field Size in Bytes}\\ \cline{2-3}
 Fields & No Profile & BLE Profiled \\ \hline
 Version & 5 & 1   \\
 Serial Number & 18 & 6  \\
 Signature & 15 & 0 \\
 Issuer & 114 & 0 \\
 Validity & 32 & 0 \\
 Subject & 168 & 0 \\
 Subject public key info & 294 & 32 \\
 Issuer and subject unique ID & 0 & 0 \\
 Extensions & 596 & 0 \\
 Signature Algorithm & 15 & 0 \\
 Signature & 261 & 64 \\ \hline
 Total & 1518 & 103
 \\ \hline
\end{tabular}}
\label{table2}
\end{table}

Figure \ref{Certificate} below shows the ECDSA signature on our BLE profiled Certificate.
\begin{figure}[h]
    \centering
    \includegraphics[keepaspectratio,width=0.9\textwidth]{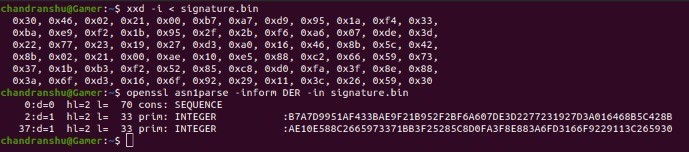}
    \caption{ECDSA Signature}
    \label{Certificate}
\end{figure}

\subsection{Cryptography Involved}
\begin{itemize}
\small
    \item ECDH: It is an key exchange algorithm, where both the client and server generates an Elliptic Curve Public/Private key pair, to establish a Shared Secret (DHKey) over an insecure channel. The DHKey is established using the secure Private Key and the Public Key transmitted using the Digital Certificate. The elliptic curve used here is NIST P-256 (prime256v1) \cite{ECC}.
    \item ECDSA: It is a variant of Digital Signature Algorithm applied to Elliptic Curves. It works on the hash of a message. ECDSA certificates are used for providing authentication/ownership of the public key. Since the security of a key is dependent on its size and the algorithm as well hence we have made use of ECDSA over RSA because ECDSA provides us the same level of security as RSA but with smaller keys. To give an example, A 256-bit elliptic curve key provides as much protection as a 3,072-bit RSA key. Smaller keys have faster algorithms for generating signatures. Smaller public keys also mean smaller certificates and less data to pass around to establish a connection. This means faster and energy efficient connections \cite{ECDSA}.
    \item AES-CMAC: It is a Message authentication code function which relies on the block cipher AES as the building block. It is used in our solution to generate confirm value and LTK. This LTK is then used for encryption/decryption purpose \cite{rfc4493}.
\end{itemize}

\section{Implementation and evaluation}

\subsection{Experiment setup}
The proposed solution has been tested on ESP-32 as well as on arduino nano 33 IoT board (having BLE v4.2) and android smartphones. We have made use of Zephyr OS and ESP-IDF, both designed for embedded and resource constrained devices. It is installed on a machine with Ubuntu 20.04.2 LTS, intel i7 (7th Gen). ESP-32 board uses Tensilica Xtensa LX6 microprocessor operating at 240 MHz clock, 520 KB SRAM, 4MB of Flash memory having Cryptographic hardware acceleration. The nano 33 IoT uses Arm Cortex M0\textsuperscript{+} 32-bit SAMD21 processor operating at 32 KB SRAM, 256 KB flash memory, 48 MHz clock speed.

\subsection{Energy Consumption}
We discuss the energy consumption by the cryptographic algorithms, which are used in our proposed model. Arduino nano 33 IoT, with Arm Cortex M0\textsuperscript{+} 32-bit SAMD21 processor running at 48 MHz was used for testing the energy consumption. The software was implemented using the cryptographic libraries available on ARM mbedTLS. The total energy consumption by the device is listed in Table \ref{energy}. The results shown in the table is the average energy consumption by the device, while performing these cryptographic algorithms. Clearly, the ECC algorithm consumes a large amount of energy compared to AES and SHA algorithm.

\begin{table}[h]\centering
\caption{Energy Consumption of Cryptographic algorithms on ARM Cortex M0+}
{\begin{tabular}{|p{7.0cm}|p{4.0cm}|} \hline
 Cryptographic Algorithm & Energy \\ \hline
 P-256 ECDHE & 34.13 mJ/Op \\
 P-256 ECDSA-Sign & 11.86 mJ/Op  \\
 P-256 ECDSA-Verify & 35.09 mJ/Op \\
 SHA-256 Message Digest & 0.041 $\mu$J/B \\
 SHA-256 HMAC (64-Byte Key) & 0.056 $\mu$J/B \\
 AES-128 & 0.134 $\mu$J/B \\ \hline
\end{tabular}}
\label{energy}
\end{table}

Peripheral devices remain in sleep mode for the majority period of time except when it has to communicate with the central (master) device. There are four different phases in a connection event: wakeup and pre-processing, transmit, receive, and post-processing. Inter-Frame Space (IFS) is a short delay time between subsequent transmission and reception.
The energy consumption by a peripheral device during transmission ($E_{T}$) and reception ($E_{R}$) is composed of : Energy to wakeup ($E_{WU}$), transmit ($E_{TX}$), receive ($E_{RX}$), waiting for IFS ($E_{IFS}$) and for post-processing and sleep ($E_{SLP}$) and is given by:\\
\begin{equation}
    \texttt{ $E_{T}=E_{WU}+nl_{HDR}E_{RX}+(2n-1)E_{IFS}+(nl_{HDR}+l_{P})E_{TX}+E_{SLP}$ }
\end{equation}

\begin{equation}
    \texttt{$E_{R}=E_{WU}+(nl_{HDR}+l_{P})E_{RX}+(2n-1)E_{IFS}+nl_{HDR}E_{TX}+E_{SLP}$}
\end{equation}

Where $l_{HDR}$ (= 14 bytes) denotes the length of BLE header and trailer, $l_{P}$ is the total payload that can fit in the link layer, and n is the number of fragments of the payload sent.

As can be seen in Table \ref{energy}, the cryptographic algorithm that accounts for the majority of energy consumption are the ECC algorithms. Now, considering that BLE profiled certificate have been used, the energy spent by the device when sending and receiving the certificates can be calculated by considering only the ECC algorithms and ignoring the AES and SHA algorithms. Therefore, total energy consumption when sending and receiving certificates can be estimated using \texttt{P-256 ECDSA-Verify} operation, which consumes \texttt{35090$\mu$J}.\texttt{P-256 ECDSA-Sign} operation, which is used only if the device sends its digital certificate to the other device signed by it's private key, it consumes \texttt{11860$\mu$J}. This is the only additional energy consumption required for our proposed model, since the rest of the steps in our model follows of the same mechanism used in LE secure connection Just works model.

\subsection{Memory usage}
Two different Certificates have been evaluated: A Regular X.509 certificate used in the Web and a BLE profiled Certificate. The total size of the certificates along with their individual field size are shown in the Table \ref{energy}. As can be clearly seen from the table, the difference between the size of both the certificates is noticeable and hence it will help us in terms of storage and energy consumption. This is achieved by keeping only the necessary details while removing those which can be tolerated. 

\subsection{Simulation for testing the security of proposed authentication method using Verifpal tool}

Through the security verification using the \texttt{Verifpal} tool, we show that the proposed scheme is secure against MITM attack and properties such as Confidentiality and Authentication is taken care of. Verifpal uses intuitive language for modelling the protocols, its internal logic is based on deconstruction and reconstruction of abstract terms, just like \texttt{ProVerif}. However, Verifpal is meant to describe the protocols close to how one may describe it in an informal conversation. Verifpal uses its own cryptographic functions and does not allow users to make use of their own cryptographic primitives. Some of them includes: ENC and DEC meaning encryption and decryption and AEAD\_ENC AND AEAD\_DEC meaning authenticated encryption and decryption. Verifpal works on analysing the model, that helps users to find which values it is able to deconstruct or reconstruct based on attacks possible in real world scenario.

The first thing we do in Verifpal is to declare an attacker. The attacker can be an active attacker or a passive attacker. We describe our attacker to be active using attacker[active].
The next step is to declare the two principals involved in our model using principal Alice and principal Bob. Next, both Alice and bob generates their private keys using knows private longtermA/longtermB and then generates their public keys using the generator and private key. The public keys are then sent to each other while keeping them in guards using Bob -$>$ Alice: [longtermPublicB], indicating that the keys are pre-authenticated using the certificate. Also, both the devices generate a random nonce value using generates Na/Nb. The generates indicates freshness of the nonce value. Finally, both parties calculate the shared secret using their private keys and each other’s public key.
At last, nonce values of both the parties are shared with each other and the long-term key (LTK) is calculated using MAC (DHKey, message) by both the parties. Here message contains 4 inputs, nonce value and public keys of both Bob and Alice. It is added using CONCAT (a, b...) operation. Finally, we specify the queries we want to test using, queries [authentication? Bob -$>$ Alice: m]. 

The simulation result is provided in the Figure \ref{fig6} below. As can be clearly seen the output shows that all the test cases are passed and verified. This verified that our proposed scheme is secure against active MITM attacks and authentication is guaranteed. Moreover, the confidentiality property is also verified along with the equivalence property of two LTK (LTK1 and LTK2).

\begin{figure}[h]
    \centering
    \includegraphics[keepaspectratio,width=0.9\textwidth]{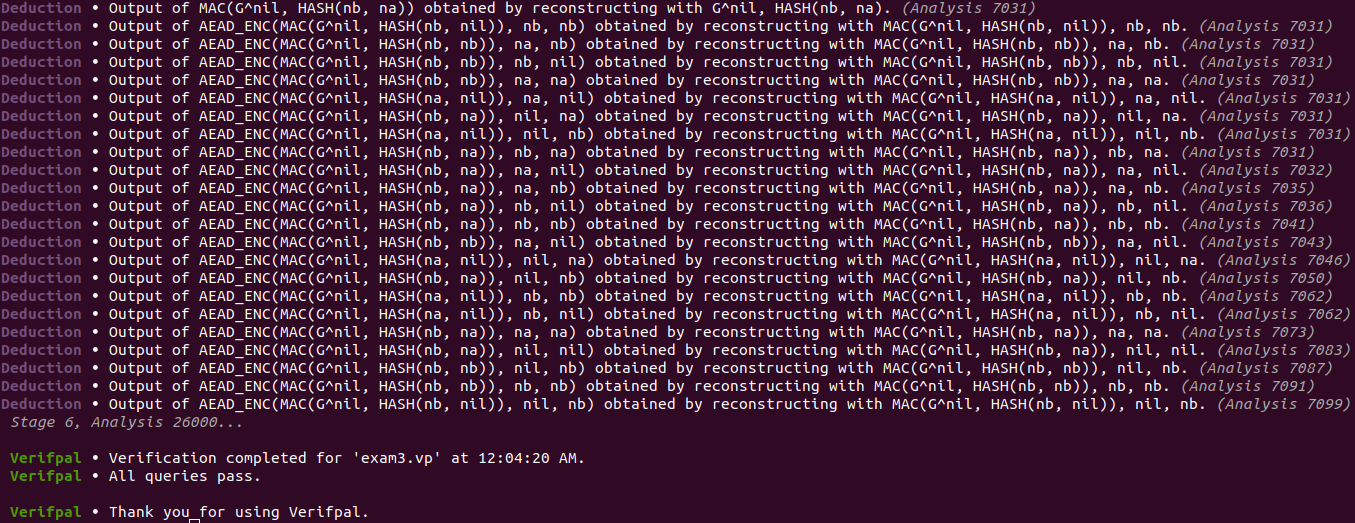}
    \caption{Verification result}
    \label{fig6}
\end{figure}



\section{Conclusion and Future Work}
In this work, we have proposed and implemented a novel secure digital certificate based authentication mechanism between two BLE devices having just works as their associative model. The solution provided has been verified using Verifpal and all the security goals are passed by our protocol. Since the proposed model is a supplement to the already existing pairing mechanism, there is no drastic change required in the Security manager protocol to implement our solution.

Also, we have proposed a novel BLE profiled certificate for authenticating the BLE devices, the profiled certificate requires significantly less memory usage (approx 93\% reduction when compared with normal X.509 certificate) and (approx 67 \% reduction when compared with IoT profiled certificate). The decrease in size of the certificate will also help in reduction of energy consumption by the device and therefore will adhere with the low energy property of BLE devices.

Although, we have proposed and tested our solution for Just works pairing model, this solution can even be incorporated to all other associative models (Passkey entry, Numeric Comparison) used in BLE, hence making a common pairing solution for any type of BLE device, resulting in less resource requirements and change in BLE architecture. Moreover, we have also tested our solution without involving any pairing mechanism, by implementing our model above GATT layer, transparent to application layer using WolfSSL. This is done for the devices that are having BLE version less than 4.2, because such devices can't make use of LE secure connections. Using this, we don't even need to change the BLE stack, because the solution works above GATT layer. This particular solution is also tested for devices with BLE v 4.2 and is compatible to work with all BLE versions.

In the future we want to work on implementing the proposed solution to many different versions of BLE devices and also on the devices that use different associative model than Just works. We also plan to provide more efficient solution for certificate revocation process, when the private key of the device is compromised.





\bibliographystyle{elsarticle-num}

\bibliography{sample}

\end{document}